\documentclass[preprint,showpacs,aps,prd,nofootinbib,floatfix,amsmath,amssymb]{revtex4-1}
\usepackage{pslatex}
\usepackage[pdftex]{graphicx}

\usepackage{epsfig}
\usepackage{color}
\usepackage{cancel}
\usepackage{slashed}
\usepackage{amssymb}
\usepackage{amsmath}
\usepackage{hyperref}

\usepackage[utf8x]{inputenc}
\usepackage{color}
\usepackage{multirow}
\usepackage{slashed}
\usepackage{ulem}
\usepackage{pstricks}
\usepackage[caption=false]{subfig}
\begin{document}

\title{On Landau pole in the minimal 3-3-1 model }


\author{Edison T. Franco}%
\email{edisonfranco@uft.edu.br}
\affiliation{
Universidade Federal do Tocantins - Campus Universit\'ario
de Aragua\'\i na\\
Av. Paraguai (esquina com Urixamas), Cimba \\ Aragua\'\i na - TO, 77824-838, Brazil
}

\author{V. Pleitez}%
\email{vicente@ift.unesp.br}
\affiliation{
Instituto  de F\'\i sica Te\'orica--Universidade Estadual Paulista (UNESP)\\
R. Dr. Bento Teobaldo Ferraz 271, Barra Funda\\ S\~ao Paulo - SP, 01140-070,
Brazil
}

\date{12/09/2017
}
\begin{abstract}
We show that in 3-3-1 models the existence of a Landau-like pole in the coupling constant related to the $U(1)_X$ factor, $g_X$, in a certain value of
$\sin^2\theta_W $, arises only assuming that the condition to match the gauge coupling constants of the standard model, $g_{2L}$, with that of the 3-3-1 model, $g_{3L}$, is valid for all energies. However, if we impose that this matching condition is valid only at a given energy, say $\mu = M_Z$, the pole arises when $\sin^2\theta_X(\mu_{LP})=1$, which is the only weak mixing angle in the models. The value of $\mu_{LP} $ depends on the energy scales, $\mu_m$ and $\mu_{331}$, in which the matching and the 3-3-1 symmetry is fully realized, respectively. We also show that $g_{2L} $ and $g_{3L} $ have different running with energy. Therefore, differently from what is usually assumed in the literature, these couplings can not be considered equal for all energies. As a consequence, the fermion couplings with neutral vector bosons are different if we write them in terms of $\sin\theta_X$ instead of $\sin\theta_W $.
\end{abstract}


\maketitle

\section{Introduction}
\label{sec:intro}


Currently there is no doubt that models with $SU(3)_C\otimes SU(3)_L\otimes U(1)_X$  gauge symmetry (3-3-1 for short) are interesting candidates for physics beyond the standard model (SM).
As in the SM, 3-3-1 models have three couplings: $g_s$ related to $SU(3)_C$, $g_{3L}$ related with $SU(3)_L$, and $g_X$ corresponding to the $U(1)_X$ factor,  and we can define an angle  as $\tan\theta_X~=~g_X/g_{3L}$. Since the early nineties, it has been known that all these models have a Landau pole in the coupling $g_X$ which occurs at an energy that depends on the representation content of the model.
In the case of the minimal 3-3-1 model (hereafter m331 for short)~\cite{Pisano:1991ee,Frampton:1992wt}, which is called minimal in the sense that only the known leptons are in triplets of $SU(3)$, $(\nu_l,l,(l^c))_L;\,l=e,\mu,\tau$, the relation
\begin{equation}
t^2\equiv \tan^2\theta_X=\frac{g^2_X}{g^2}=\frac{s^2_W}{1-4s^2_W},
\label{pole}
\end{equation}
($s^2_W\equiv \sin^2\theta_W$) implies that the pole accurs when $s^2_W(\mu_{LP})=1/4$, where $\mu_{LP}$ is the energy scale at which the coupling $g_X$ goes to infinity, and the model becomes non-perturbative. Eq.~(\ref{pole}) is also valid in the 3-3-1 model with heavy leptons (331HL for short) in which the lepton sector includes heavy charged partners $(\nu_l,l,E^+)_L;\,l=e,\mu,\tau$~\cite{Pleitez:1992xh}.

The existence of a Landau-like pole at a given energy $\mu_{LP}$ is more restrictive in the models of Refs.~\cite{Pisano:1991ee,Frampton:1992wt,Pleitez:1992xh}, since it has been shown that $s^2_W(\mu_{LP})=1/4$~\cite{Ng:1992st} at energies of a few TeVs, i.e., $\mu_{LP}\approx 4$ TeV~\cite{Dias:2004dc}. However, such a pole does exist but not at a given value of $s^2_W$ but in the angle that appears in the model, $s^2_X\equiv \sin^2\theta_X$.

Here we will show that the existence of the Landau pole in $g_X$ when $s^2_W=1/4$, is due to the imposition of the matching conditions $g_{3L}=g_{2L}$ and that in Eq.~(\ref{pole}) are valid for all energies.
On the other hand, we will assume that these conditions are valid only at a given energy, say $\mu=\mu_{matching}\equiv \mu_m$. In this energy scale, we change the symmetry of SM to symmetry 3-3-1, but only considering the degrees of freedom that are common to SM. The Landau pole in $g_X$ occurs when $s^2_X=1$ at greater energies than it has been thought before.

The outline of this paper is as follows. In Sec.~\ref{sec:lpole} we review the origin of the Landau pole in the coupling $g_X$ in the context of the m331 model. In Sec.~\ref{sec:running} we calculate the running of $s^2_X$ using the renormalization group equations given in the Appendix~\ref{sec:rge}. In Sec.~\ref{sec:nc} we calculate the lepton neutral couplings with $Z_1$ and $Z_2$.
The Sec.~\ref{sec:con} is devoted to our conclusions.

\section{The origin of the Landau pole}
\label{sec:lpole}

Let us first review the origin of the Landau pole in the 3-3-1 models considered here.
In the m331 it was shown that the electric charge $e$ is related with the coupling $g_{3L}$ of $SU(3)_L$ and $g_X$ of $U(1)_X$ as follows~\cite{Pisano:1991ee}
\begin{equation}
e=g_{3L}\frac{\sin\theta_X}{\sqrt{1+3\sin^2\theta_X}}\;\;(a),\quad e=g_X\frac{\cos\theta_X}{\sqrt{1+3\sin^2\theta_X}}\;\;(b).
\label{ecdef}
\end{equation}
The relations in Eq.~(\ref{ecdef}) substitute $e=g_{L}s_W$ and $e=g_Yc_W$ in the SM, respectively, but notice that there is no $\theta_W$ angle in the model, only $\theta_X$.

Assuming that $g_{3L}(\mu_m)=g_{2L}(\mu_m)$, and the condition
\begin{equation}
\left.\frac{\sin^2\theta_X(\mu)}{1+3\sin^2\theta_X(\mu)}\right\vert_{\mu=\mu_m}=\sin^2\theta_W(\mu_m),
\label{mc}
\end{equation}
we obtain $s^2_X=s^2_W/(1-3s^3_W)$ and $c^2_X~=~(1~-~4s^2_W)/(1-3s^2_W)$, then Eq.~(\ref{pole}) follows which implies the existence of a Landau-like pole in $g_X$ when $s^2_W~=~1/4$.
If we use $\mu_m=M_Z$, from Eq.~(\ref{mc}) we get the value $\sin^2\theta_X(M_Z)\approx 0.7555$ [$\theta_X(M_Z)\approx 60^o$], where we have used the $s^2_W~=~0.23129$~\cite{Olive:2016xmw}, choosing the central values.

The pole also arises as follows: using the relation $1/e^2~=~4/g^2_{3L}+1/g^2_X$ among the electric charge and the coupling constants $g_{3L}$ and $g_X$, the matching condition with the SM is now introduced by defining $1/g^2_Y=3/g^2_{3L}+1/g^2_X$ and assuming $g_{3L}=g_{2L}\equiv g$, and using $\tan\theta_W=g_Y/g$ we obtain  the relation in Eq.~(\ref{pole})~\cite{Ng:1992st}.

However, in the context of the m331 model only the matching conditions with the electric charge in Eq.~(\ref{ecdef}) does exist, and we can rewritten them as follows:
\begin{equation}
\frac{\alpha}{\alpha_{3L}}=\frac{s^2_X}{1+3s^2_X},\quad \frac{\alpha_X}{\alpha}=\frac{1+3s^2_X}{c^2_X}.
\label{pole2}
\end{equation}
Notice from these equations that, when $s^2_X\to 1$ then $c_X\to 0$ i.e., $\alpha_X\to \infty$. Hence, the Landau-like pole arises now in the angle $\theta_X$, which is in fact the only electroweak angle in this model.

\section{The running of $s^2_X$}
\label{sec:running}

Firstly, let us show that in general $\alpha_{3L}(\mu)\not=\alpha_{2L}(\mu)$, if $\mu\not=\mu_m$. This is expected because both models have different degrees of freedom embedded into different gauge symmetries.
The running of the inverse of coupling constants are obtained using the renormalization group equations in Eq.~(\ref{aa2}), where $\mu$ is in the interval $\mu\in [M_{Z},10\, \textrm{TeV}]$. Recall that the $\mu_m$ is the energy at which $\alpha^{-1}_{2L}(\mu_m)=\alpha^{-1}_{3L}(\mu_m)$ and the SM symmetry changes to 3-3-1 symmetry but with the degrees of freedom that are commom to the SM. We introduce the energy scale $\mu_{331}$ at which not only the symmetry is the 3-3-1 but all the degrees of freedom of the m331 model are used. Although these scales are in general different, in order to see how $\alpha^{-1}_{2L}$ runs different than $\alpha^{-1}_{3L}$ we assume that $\mu_m=\mu_{331}$
and test several values for $\mu_m= M_Z$, $1.0,4.0$ TeV.

\begin{figure}
\begin{center}
\center
	\includegraphics[height=0.27\columnwidth]{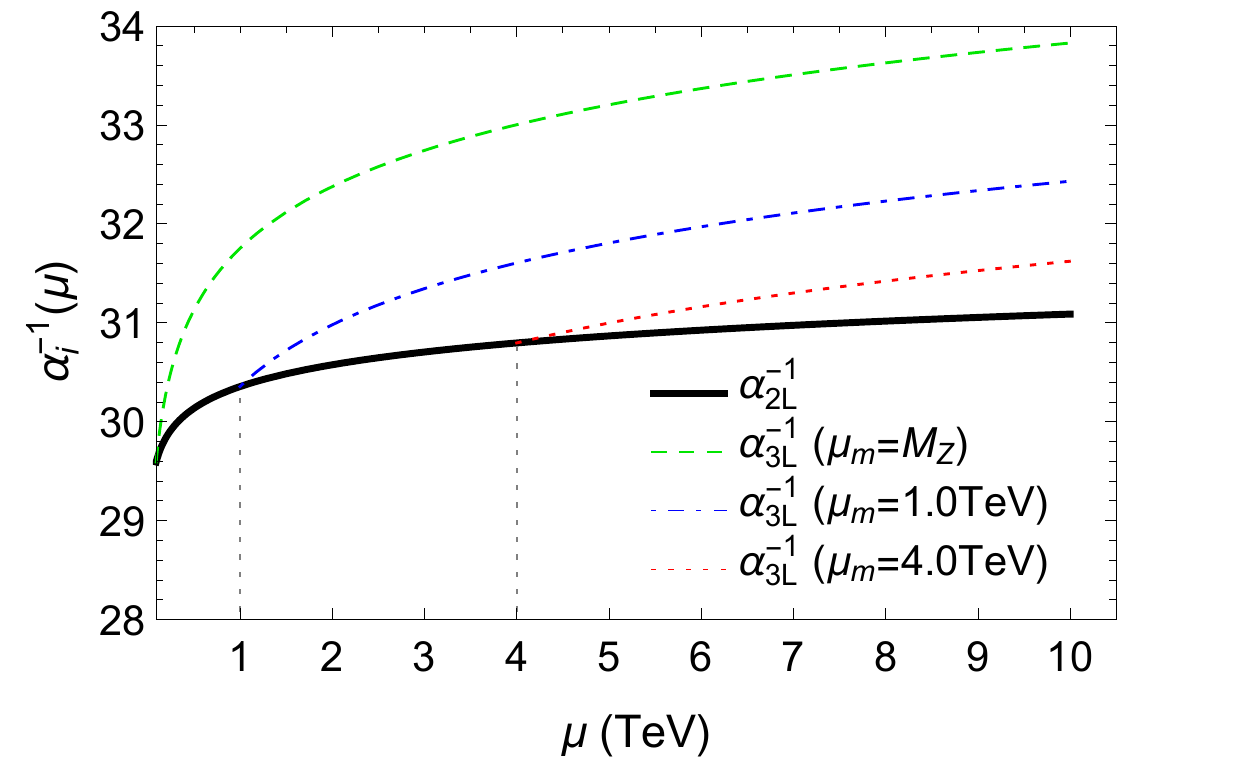}
\hfill
		\caption{ \label{fig:fig1}
The runninig of $\alpha^{-1}_{2L}$ and $\alpha^{-1}_{3L}$ using $\mu_m=\mu_{331}$ and three values: $\mu_{331}=M_Z$,1.0 TeV and 4.0 TeV.
}
	\end{center}
\end{figure}

In order to obtain $\alpha^{-1}_{2L}$ we use the $b_i$ coefficients given in Eq.~(\ref{aa3}) i.e.,
considering only the SM symmetry and its respective degrees of freedom.
In Fig.~\ref{fig:fig1} the running of $\alpha^{-1}_{2L}$ begining at $\mu=M_Z$ is shown by the continuos (black) line. When $\mu>\mu_m$ the gauge symmetry is changed to 3-3-1. For $\alpha^{-1}_{3L}$ we start from $\mu=\mu_m$ using the coeficients in Eq.~(\ref{aa6}) and the initial condition $\alpha^{-1 }_{2L}(\mu_m)=\alpha^{-1 }_{3L}(\mu_m)$.

We have considered three  different values of $\mu_m$ and shown the results in Fig.~\ref{fig:fig1}:
(i) $\mu_m=M_Z$, and the running of $\alpha^{-1}_{3L}$ is shown by the (green) long-dashed line;
(ii) $\mu_m=$1.0 TeV, this results in the $\alpha^{-1}_{3L}$ given by the (blue) dot-dashed line;
(iii) $\mu_m=4.0$ TeV, and the running of $\alpha^{-1}_{3L}$ is shown by the (red) dotted line. We see from Fig.~\ref{fig:fig1} that in general $\alpha_{2L}\not=\alpha_{3L}$.

Next, we calculate the energy at which the Landau pole arises when $s^2_X(\mu_{LP})=1$ and using the initial value $s^2_X(M_Z)=0.7555$.
As discussed above, $\mu_m$ is the energy at which $\alpha^{-1}_{2L}=\alpha^{-1}_{3L}$ and the condition in Eq.~(\ref{mc}) is valid. In Figs.~\ref{fig:dois} we have chosen $\mu_m=M_Z$. This is justifiable once the m331 model, as any other extension of the SM, must satisfy the predictions of the SM in some but not in all the energies. Since the most accurate measurements confirming the predictions of the SM are made at the Z pole, we consider this is the scale in which the two models coincide in their predictions. Hence, $\mu_m\not=\mu_{331}$ and, as above,  $\mu_{331}$ denotes the energy at which the full 331 symmetry and all the degrees of freedom of the model are taken into account.

Firstly, considering only the SM-like degrees of freedom and symmetry and using $s^2_W(\mu)=1/(1+\alpha_2/\alpha_Y)$ and the matching in  Eq.~(\ref{mc}), we obtain:
\begin{equation}
s^2_X(\mu)=\frac{1}{\frac{\alpha_{2L}}{\alpha_Y}-2}.
\label{run1}
\end{equation}

The running of $s^2_X$ given by Eq.~(\ref{run1}), using the coefficients in Eq.~(\ref{aa3}), is shown by the thick continuous (black) curve in Fig.~\ref{fig:dois}. Notice that in this case the Landau pole arises at an energy of $\mu_{LP}\approx3.6$ TeV. This value is near the value obtained in Ref.~\cite{Dias:2004dc}.

Secondly, when the symmetry is the 3-3-1 we use
\begin{equation}
s^2_X(\mu)=\frac{1}{1+\frac{\alpha_{3L}}{\alpha_X}},
\label{run2}
\end{equation}
and proceed as follows.
When $\mu_m\leq\mu\leq\mu_{331}$ we use the $b_i$ coefficients given by Eq.~(\ref{aa4}), when only the triplet $\eta$ is considered, and the coefficients given by Eq.~(\ref{aa5}), when both triplets $\eta$ and $\rho$ are taken into account. In these cases, which we consider as an interemediate stage, the running of $s^2_X$ in Eq.~(\ref{run2}) is shown by the lower thin (black) curves in Figs.~\ref{2a} and \ref{2b}, using Eqs.~(\ref{aa4}) and (\ref{aa5}), respectively.
In the first case the Landau pole would arise at 333.4 TeV, and in the second case at 124.6 TeV.
These lower thin (black) curves will be used for comparing with the subsequent cases: when $\mu>\mu_{331}$ we will consider the full 3-3-1 symmetry and all its degrees of freedom, given by Eq.~(\ref{aa1}) and the coefficients given by Eq.~(\ref{aa6}).

The latter cases are shown in Fig.~\ref{fig:dois} for several values of $\mu_{331}$ (the numbers inside parenthesis refer to Fig.~\ref{2b}).
We consider $\mu_{331}=M_Z$, shown by the dashed (green) curve with $\mu_{LP}=7.7$ TeV in both cases. When $\mu_{331}=0.3$ TeV it is shown by the dot-dashed (blue) curve and $\mu_{LP}=13.3(12.2)$ TeV, when $\mu_{331}=1.0$ TeV it is shown by dotted (red) curve with $\mu_{LP}=23.2(19.4)$ TeV, and finally with $\mu_{331}=4.0$ TeV, appearing in the dot-dot-dashed (pink) curve with $\mu_P=43.8(33.1)$ TeV. We have also calculated, without showing in the Fig.~\ref{fig:dois}, for $\mu_{331}=14$ TeV we obtain $\mu_{LP}=77.8(53.7)$ TeV; and for $\mu_{331}=20.0$ TeV,  $\mu_{LP}\simeq 91.6(61.6)$ TeV is obtained.

Since, in general, the energy scale of the full 3-3-1 symmetry is different from the gauge couplings matching scale, we have verified that if $\mu_m$ is larger (smaller) than $M_Z$ the pole occurs at lower (higher) energies than those shown in Figs.~\ref{fig:dois}.

It is interesting to note that in the m331 model, at least in the lepton sector which involves only the known leptons, the 3-3-1 symmetry could completely replace the SM symmetry. In this case, the Landau pole is reached at energies greater than those obtained here. On the other hand, the energy scale $\mu_{331} $ should be near the mass of $Z_2$. See below.

\section{Neutral currents}
\label{sec:nc}

In the previous section we have shown that there is no \textit{ a priori} a fixed energy scale at which the Landau-like pole arises. It depends on the energy at which the new degrees of freedom of the model are excited.

If the weak mixing angle in the model is $\theta_X$ it is $s^2_X$, and  not $s^2_W$, which must appear in the neutral current couplings, $\tilde{g}^i_{V,A}$. For instance, let us consider the neutrino and charged leptons couplings with $Z_1\approx Z$ and $Z_2\approx Z^\prime$ which in general are parametrized as
\begin{equation}
\mathcal{L}^{NC}_{331}=-\frac{g_{3L}}{2}\sum_i\bar{\psi}_i\gamma^\mu[(\tilde{g}^i_V-\gamma_5\tilde{g}^i_A)Z_{1\mu}
+(f^i_V-\gamma_5f^i_A)Z_{2\mu}]\psi_i,
\label{nc1}
\end{equation}
with $\psi_i=l_i,\nu_i,u_i,d_i$ and the couplings $\tilde{g}^i_{V,A}$ and $f^i_{V,A}$ being usally written in terms of the weak mixing angle $s^2_W$.

The full analytical couplings are given, without introducing $s_W$, by
\begin{eqnarray}
&&g^\nu_V=g^\nu_A=-\frac{1}{3}N_1[1-6A(1+B)-3\bar{v}^2_\rho+4\bar{v}^2_W],\nonumber \\&&
g^l_V=-N_1(1-\bar{v}^ 2_\rho),\;\;g^l_A=\frac{1}{3}N_1[1-6A(1+B)-3\bar{v}^2_W],
\label{gvga}
\end{eqnarray}
for the lepton couplings with $Z_1$, and
\begin{eqnarray}
&&f^\nu_V=f^\nu_A=-\frac{1}{3}N_2[1-6A(1-B)-3\bar{v}^2_\rho+4\bar{v}^2_W],\nonumber \\&&
f^l_V=-N_2(1-\bar{v}^ 2_\rho),\;f^l_A\!\!=\!\!\frac{1}{3}N_2[1-6A(1-B) -3\bar{v}^2_W+4\bar{v}^2_\rho].
\label{fvfa}
\end{eqnarray}
for the lepton couplings with $Z_2$. Where $\bar{v}^2_x=v^2_x/v^2_\chi$ with $v^2_x=v^2_\rho,v ^2_W$,  and $v^2_W=v^2_\eta+v^2_\rho+v^2_S$, being $v^2_W=(246\,\textrm{GeV})^2$, and $v_\eta,v_\rho,v_\chi$ are the vaccum expectation values for the triplets and $v_S$ for the sextet that are present in the model. Here we have defined
\begin{eqnarray}
&& N^{-2}_1=3\left[2(A(1+B)+\bar{v}^2_\rho-\frac{4}{3}\bar{v}^2_W-\frac{1}{3}\right]^2+(\bar{v}^2_\rho-1)^2
\frac{1+3s^2_X}{c^2_X},\nonumber\\&&
N^{-2}_2=3\left[2A(1-B)+\bar{v}^2_\rho-\frac{4}{3}\bar{v}^2_W-\frac{1}{3}\right]^2+(\bar{v}^2_\rho-1)^2
\frac{1+3s^2_X}{c^2_X},
\label{def1}
\end{eqnarray}
and
\begin{eqnarray}
&& 3A=1+\bar{v}^2_W+3\left(1+\bar{v}^2_\rho\right)\frac{s^2_X}{c^2_X},\nonumber \\&&
3A^2B^2=3A^2-\left[(1+\bar{v}^2_\rho)\bar{v}^2_W-\bar{v}^4_\rho\right]\cdot\frac{1+3s^2_X}{c^2_X},
\label{def2}
\end{eqnarray}
with $B>0$.
Notice that in the m331 model in the lepton sector there is no flavor changing neutral currents (FCNCs) mediated by the vectors $Z_{1,2}$. This is not the case in the quark sector. However, in the lepton and quark sectors, there are FCNCs mediated by neutral scalars.

In Fig.~\ref{fig:tres} we show, as functions of $v_\chi$ and for several values of $v_\rho$, the lepton neutral couplings defined in Eq.~(\ref{gvga}). Since in the Lagrangian in Eq.~(\ref{nc1}) does not appear $c_W$, and $\tilde{g}_{V,A}$ are functions that are not equal to the SM function, in Fig.~\ref{3a} we plot $\tilde{g}^\nu_{V,A}c_W\equiv G^\nu_{V,A}$, and $\tilde{g}^l_Ac_W\equiv G^l_V$ and $\tilde{g}^l_Vc_W\equiv G^l_A$ in Figs.~\ref{3b} and \ref{3c}, respectively,
which are the couplings that have to be compared at the $Z$-pole.
From Fig.~\ref{fig:tres} it is clear that for values of $v_\rho$ near 54 GeV give the values of these parameters are compatible with their measured values, even at tree level. This result was predicted in Ref.~\cite{Dias:2006ns}. In this case, the radiative corrections in the context of the m331 model must be small since  the values of these parameters are already consistent with the experimental measurements at the $Z$-pole. Other values of $v_\rho$ are also allowed but in those cases, as in the SM, radiative corrections are necessary to fit the experimental values.
In Fig.~\ref{fig:quatro} we show the neutral lepton couplings with $Z_2$:  $f^\nu_{V,A}c_W\equiv F^\nu_{V,A}$, and $f^l_Ac_W\equiv F^l_V$ and $f^l_Vc_W\equiv F^l_A$, in Figs.~\ref{4a},\ref{4b} and \ref{4c}, respectively.

At the $Z$-pole $\tilde{g}^i_{V, A} $ and $f^i_ {V, A} $ written in terms of $s^2_X $ as in
Eqs.~(\ref{gvga}) and (\ref{fvfa}) or written in terms of $s^2_W$ as in Ref.~\cite{Dias:2006ns} have the same values. However, when we consider the running of these couplings with energy it is possible to notice differences when we use $s^2_X$ instead of $s^2_W$. This is shown in Figs.~\ref{fig:cinco}. The neutral couplings in
Ref.~\cite{Dias:2006ns} are indicated in that figure as ``closing'' while those in Eqs.~(\ref{gvga}) and (\ref{fvfa}) are calculated for the same values of $\mu_{331} $ as in Fig.~\ref{2a} i.e., considering only one scalar triplet, $\eta$.
The curves in Figs.~\ref{5a} and \ref{5b} actually end at the upper end, which is the energy at which the Landau pole is reached. That is, for larger energies the neutral couplings, $G^i_{V,A}$, remain constant when the energy $\mu$ is larger than $\mu_{LP}$, at least from the perturbative point of view. For example, the curve "closing" in Fig.~\ref{5a} indicates that the couplings $G^\nu_{V,A}$ has no meaning for energies larger than 3.6 TeV, at least perturbatively, and if $\mu_{331}=4.0$ TeV, the dot-dot-dashe (pink) curve $G^\nu_{V,A}$ has no meaning only after 43.8 TeV.

Next, let us consider the neutral vector boson masses, here denoted as $Z_1$ and $Z_2$. They are given by~\cite{Dias:2006ns}
\begin{equation}
M^2_{Z_1}=\frac{g^2v^2_\chi}{2}\, A(1-B),\quad M^2_{Z_2}=\frac{g^2v^2_\chi}{2}\, A(1+B).
\label{mz1z2}
\end{equation}

In Figs.~\ref{6a} and \ref{6b} we show $M_{Z_1}$ and $M_{Z_2}$, respectively, as a function of $v_\chi$. Notice from Fig.~\ref{6a} that $M_{Z_1}$ strongly depends on the values of $v_\rho$ and that near 54 GeV,  this mass is, for all practical purposes, equal to $M_Z$ of the SM, independently of $v_\chi$. On the other hand, from Fig.~\ref{6b} we see that $M_{Z_2}$ is almost independent of $v_\rho$, but depends strongly on $v_\chi$. This is interesting because as is shown in Figs.~\ref{fig:tres}, in the m331 model, if $v_\rho\approx54$ GeV, the lepton neutral couplings to $Z_1$ are, at tree level, compatible with the observed values. It means that in the model radiative corrections should be cancel out at least at lower order for values of $v_\rho\approx54$ GeV. From the figure we obtain for $v_\chi=1.0$ TeV, $M_{Z_2}=1.2$ TeV, for $v_\chi=14$ TeV, $M_{Z_2}=16.9$ TeV. In general, $M_{Z_2}\approx 1.2\,v_\chi$.

\section{Conclusions}
\label{sec:con}

The main result of this paper is that, \textit{if} the gauge couplings matching scale is equal to $M_Z$, then the Landau pole may be in the range $3.6\leq \mu_{LP}\leq 333.4$ TeV.
Thus, the Landau pole energy obtained in \cite{Dias:2004dc} was in fact a lower limit for such an energy.
The study may be improved by considering the different particle thresholds below and above the $M_Z$ scale.
The crucial point is when the use of the full 3-3-1 symmetry is considered. In the limit in which the m331 (in which the leptons are the same as the SM) replaces completely the SM at low energies, there is no matching condition, both models coincide numerically only at the $Z$-pole.

The phenomenology of the neutral currents in the m331 has been considered in literature. For instance see Refs.~\cite{Dias:2006ns,Montero:1992jk,CarcamoHernandez:2005ka,Promberger:2007py,Buras:2014yna}. However until now
$t^2$, as defined in Eq.~(\ref{pole}), has been used in such a way that neutral current couplings, $\tilde{g}$'s and $f$' in Eq.~(\ref{nc1}), depend only on $s^2_W$ since the matching in Eq.~(\ref{mc}) is supposed to be valid at any energy.
Far from this energy, the running of $s^2_X(\mu)$ has to be taken into account. The complete set of neutral current couplings and the phenomenological consequences in this and other 3-3-1 models will be given elsewhere.

The existence of Landau pole just indicate that the theory is non perturbative for energies larger than of the pole and, for these energies, new degrees of freedom has to be taken into account. In the present case they could be new stable exotic resonances as those considered in Ref.~\cite{DeConto:2017aob}.  Another possibility, in fact much more interesting, is that before reaching the Landau pole the symmetry is enlarger and new gauge bosons arise preventing the arrival at the pole. For example $[SU(3)]^3$ in which $U(1)_X\to SU(3)_X$; or we can add new scalars or fermions in the adjunt representation to the minimal representation content of the models.

Finally, we stress that, as can be see from Figs.~\ref{3a} -- \ref{3c}, if $v_\rho\simeq54$ GeV and  $v_\chi$ is larger than 1.6 TeV, the m331 model at tree level, could never had been distinguished from the SM at LEP.

\acknowledgments
ETF thanks to IFT-UNESP for its kind hospitality,  VP thanks to CNPq for partial financial support. Both authors are
thankful for the support of FAPESP funding Grant No. 2014/19164-6.

\appendix

\section{Particle content and beta coefficients}
\label{sec:rge}

In this appendix we show the representation content of the m331 model and the $\beta $%
-coefficients for the particle content relevant for our analysis.

The representation content of the m331 is as follows: three left-handed lepton triplets, $\psi_i$, two left-handed quark anti-triplets $Q_a$, one left-handed quark triplet $Q_3$, six singlets, three right-handed $u$-type and three right-handed $d$-type, plus three right-handed quarks $j_a$ and $J_R$. Three scalar triplets $\eta,\rho,$ and $\chi$, and a scalar sextet $S$. The gauge bosons are $W^\pm, V^\pm,U^{\pm\pm}, A, Z, Z^\prime,$ and gluons. Under the 3-3-1 symmetry fermions and scalar transform as follows:
\begin{eqnarray}
&& \psi_i\sim (1,3,0),\; Q_a\sim(3,3^*,-1/3), \;Q_3\sim(3,3,2/3),\;u_{iR}\sim(3,1,2/3),\; d_{iR}\sim(3,1,-1/3),\nonumber \\&& j_{aR}\sim(3,1,-4/3),\; J_R\sim(3,1,+5/3),\quad i=1,2,3;\;a=1,2,\nonumber \\&&
\eta\sim(1,3,0),\; \rho\sim(1,3,+1),\;\chi\sim(1,3,-1),\; S\sim(1,6,0).
\label{aa1}
\end{eqnarray}

The
running of $\alpha _{i}^{-1}$ coupling constants at one loop level is given
by%
\begin{equation}
\frac{\partial \alpha _{i}^{-1}\left( \mu \right) }{\partial \mu }=-\frac{1}{%
	2\pi }b_{i}.
\label{aa2}
\end{equation}

For the SM symmetry ($i=U(1)_{Y},SU(2)_{L},SU(3)_{C}$) and for the full SM
model content: 3 quarks doublets $Q_L$, six  right-handed singlets $u$ and $d$; gauge bosons $W^\pm$, $A$, $Z$ and gluons; one scalar doublet, say $\eta$. Hence, we have as usual
\begin{equation}
b^{SM}_{i}=\left( \frac{41}{6},-\frac{19}{6},-7\right) .
\label{aa3}
\end{equation}

\begin{figure}[h!]
	\subfloat[\label{2a}]{\includegraphics[height=0.27\columnwidth]{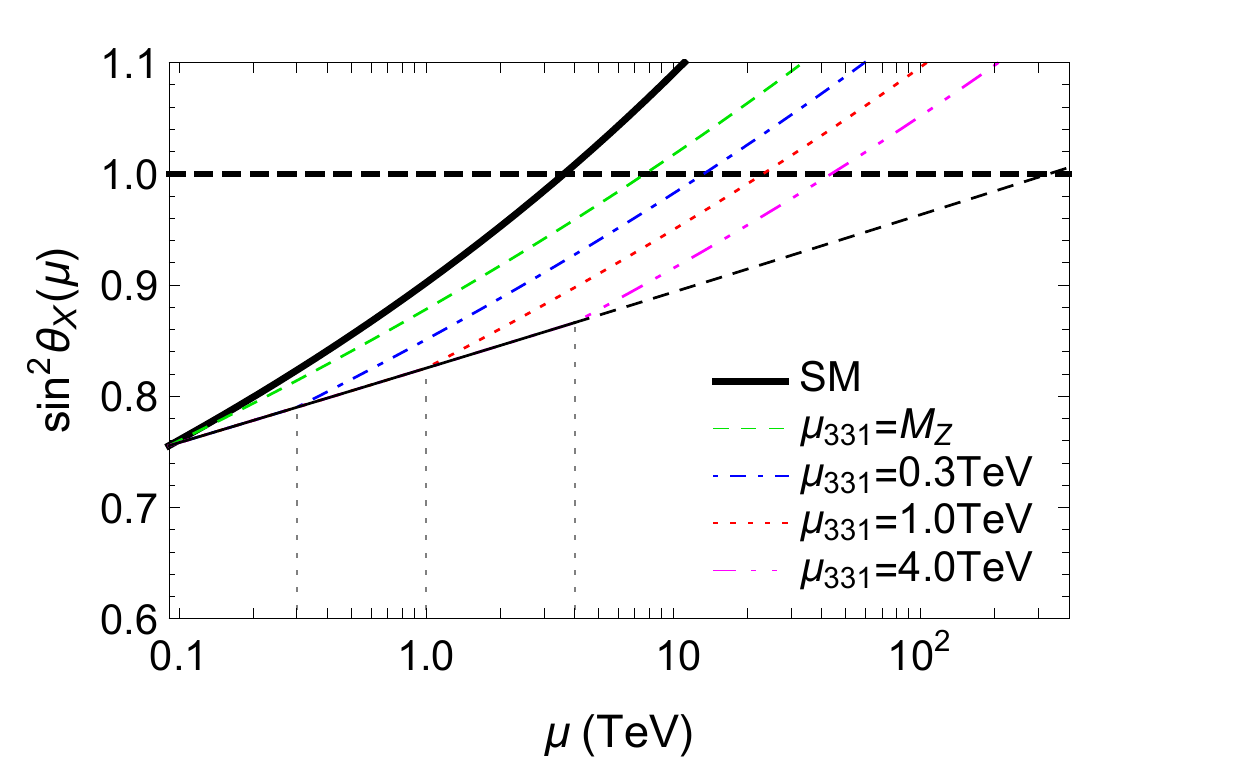}}
	\hfill
	\subfloat[\label{2b}]{\includegraphics[height=0.27\columnwidth]{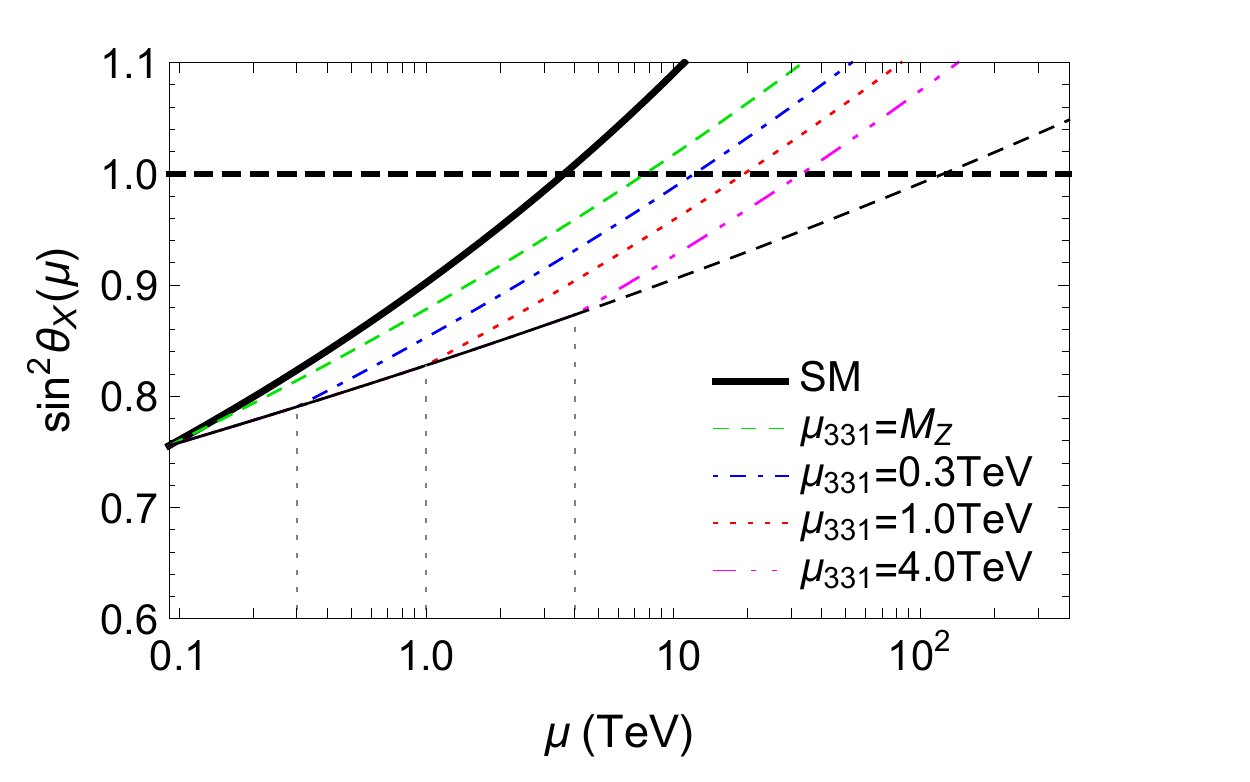}}
	\caption{The Landau pole arising when $s^2_X=1$. In all cases we are using $\mu_m=M_Z$ and we are considering the full 3-3-1 symmetry when $\mu>\mu_{331}$. In Fig.~\ref{2a} we include only the triplet $\eta$ and in Fig.~\ref{2b} the $\eta$ and $\rho$ triplets. }
	\label{fig:dois}
\end{figure}

\begin{figure}[h!]
	\subfloat[\label{3a}]{\includegraphics[height=0.27\columnwidth]{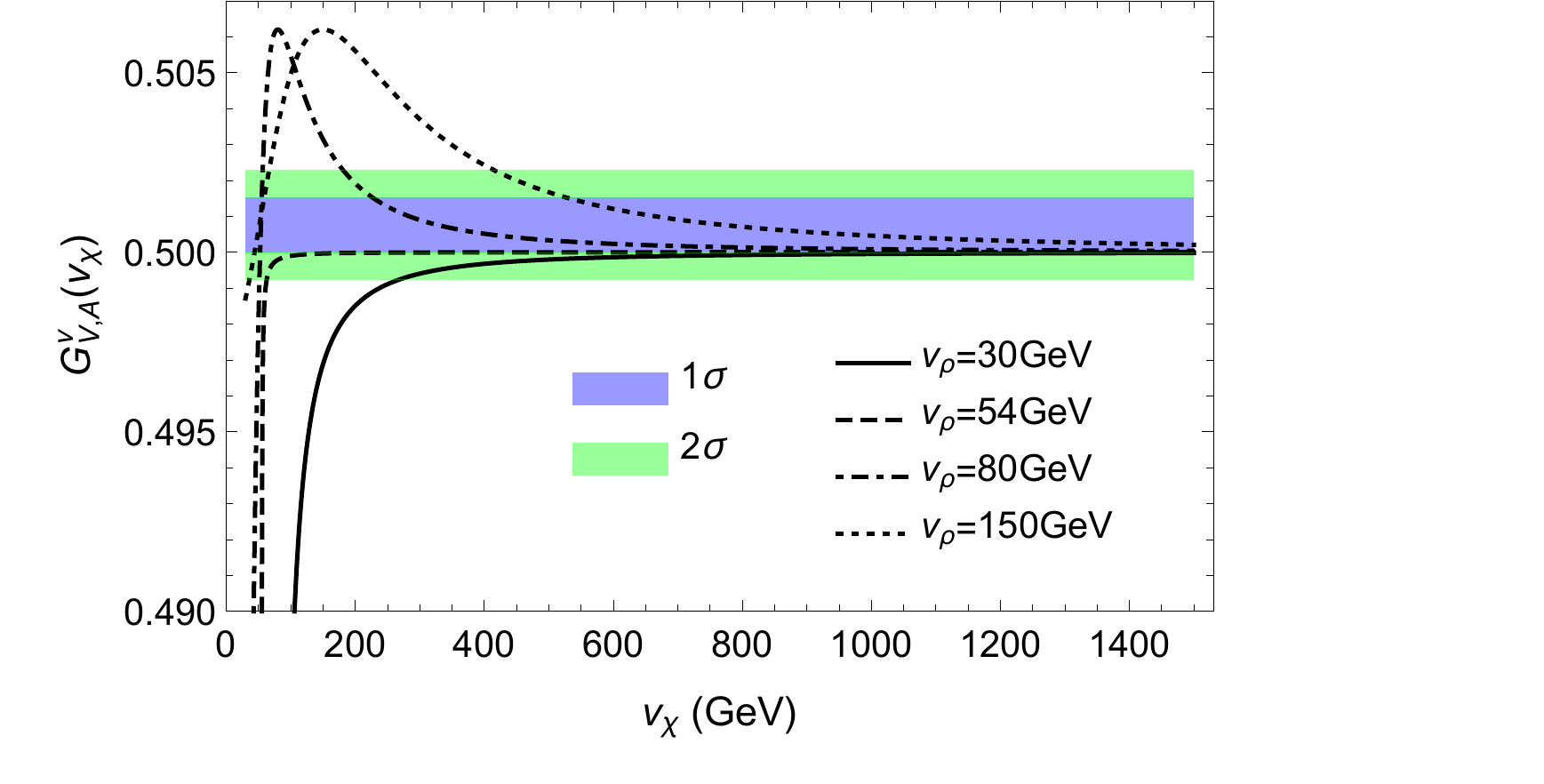}}
	\hfill
	\subfloat[\label{3b}]{\includegraphics[height=0.27\columnwidth]{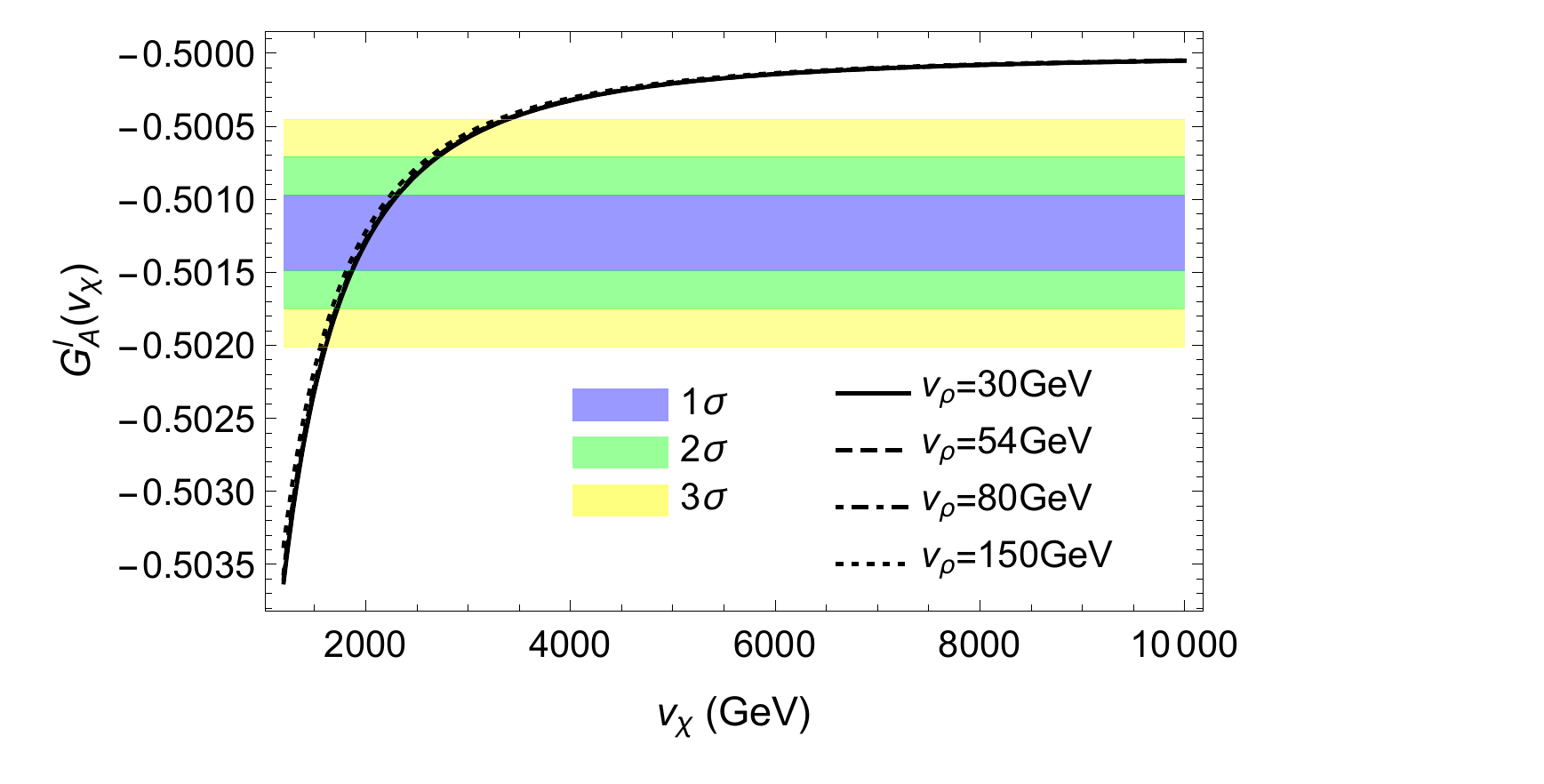}}
	\hfill
	\subfloat[\label{3c}]{
		\includegraphics[height=0.27\columnwidth]{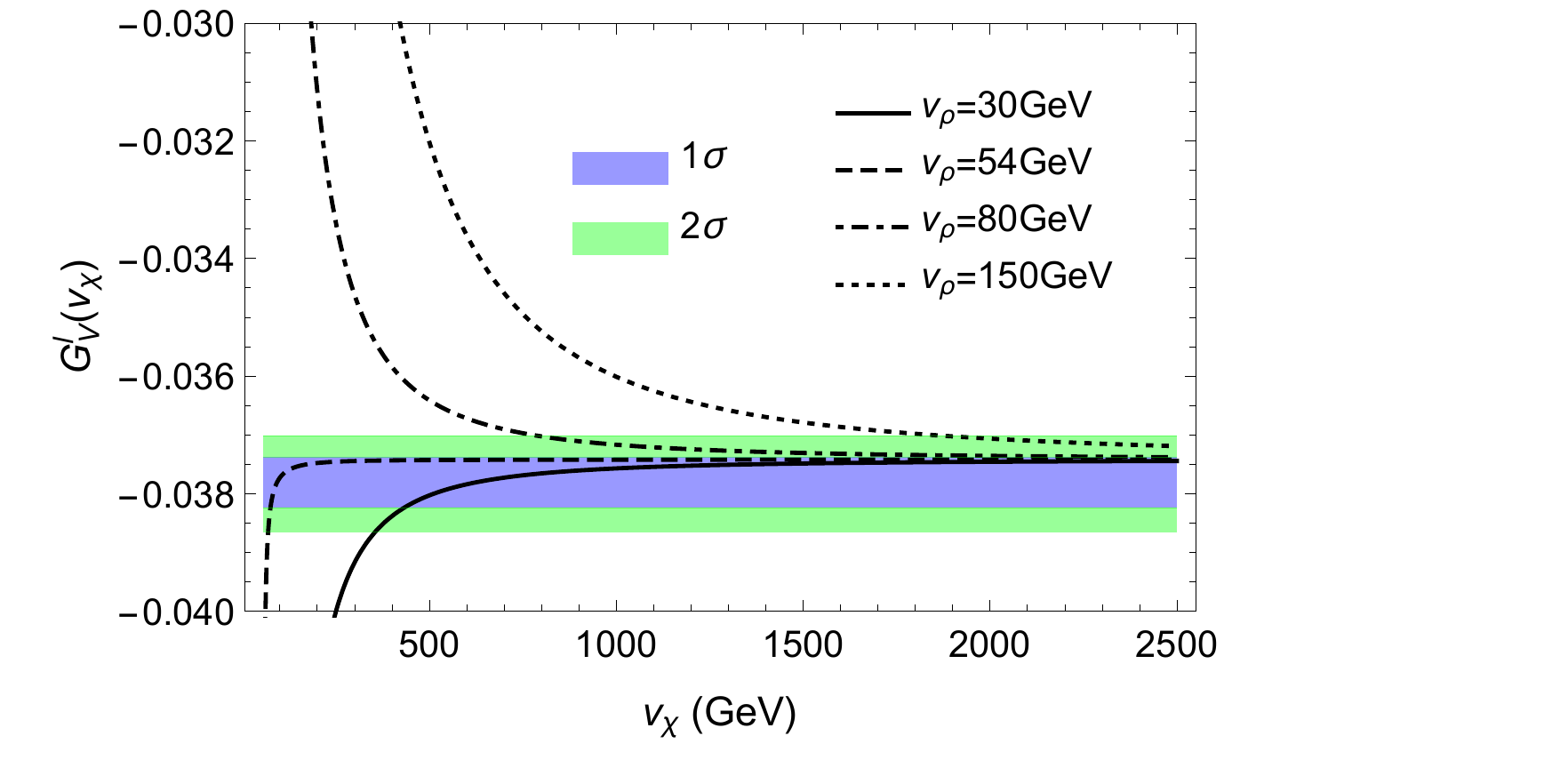}}
	\caption{Lepton neutral current couplings with $Z_1$ as a function of $v_\chi$ and $v_\rho$ at Z-pole as follows: (a) the neutrino coupling, $G^\nu_{V,A}$; (b) the charged lepton axial couplings, $G^l_A$; (c) the charged leptons vector coupling, $G^l_V$. We have define $\tilde{g}^i_{V,A}c_W\equiv G^i_{V,A}$. In all case we have shown the measurements from LEP with experimental error bands.}
	\label{fig:tres}
\end{figure}

\begin{figure}[h!]
	\subfloat[\label{4a}]{\includegraphics[height=0.27\columnwidth]{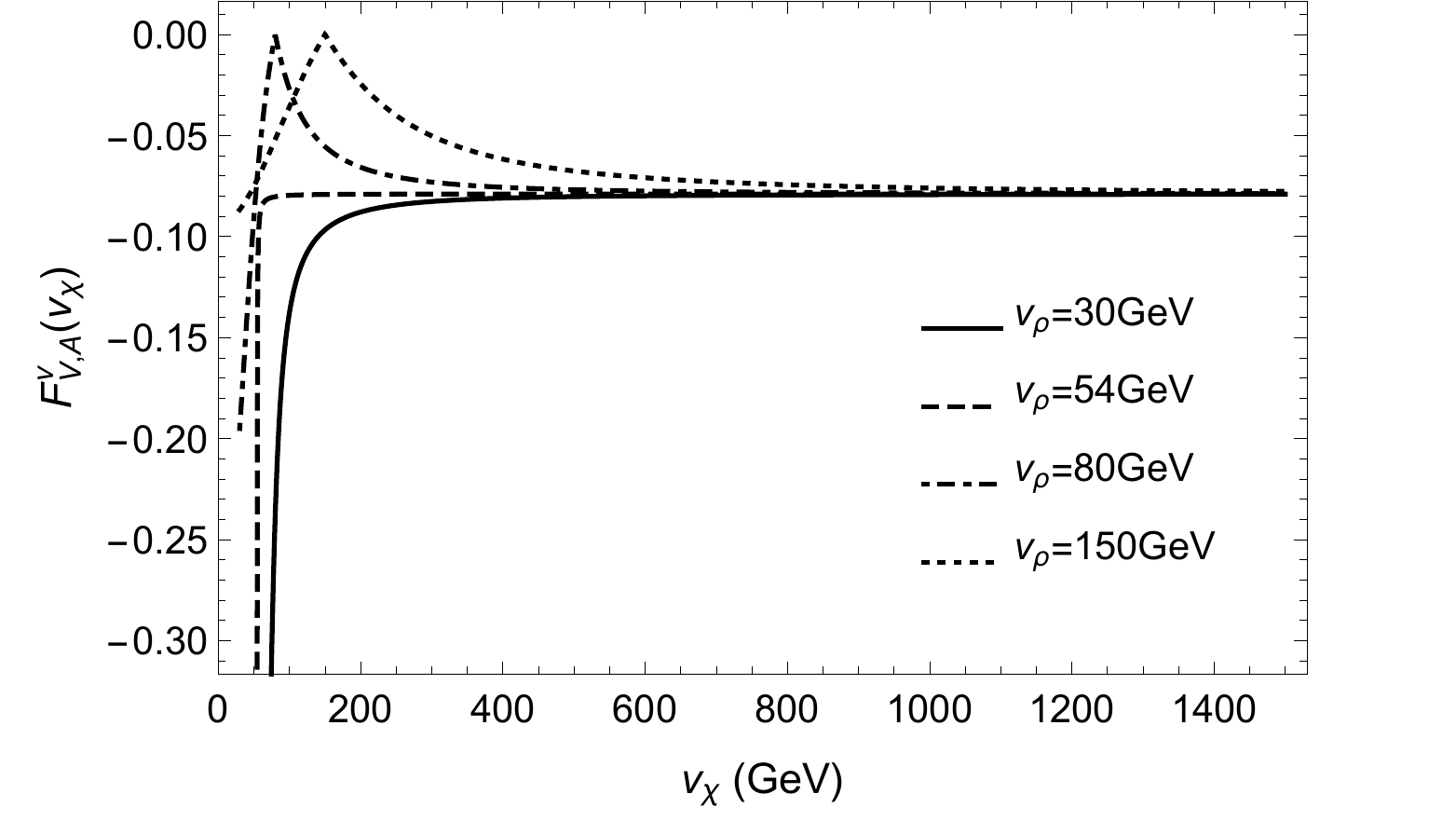}}
	\hfill
	\subfloat[\label{4b}]{\includegraphics[height=0.27\columnwidth]{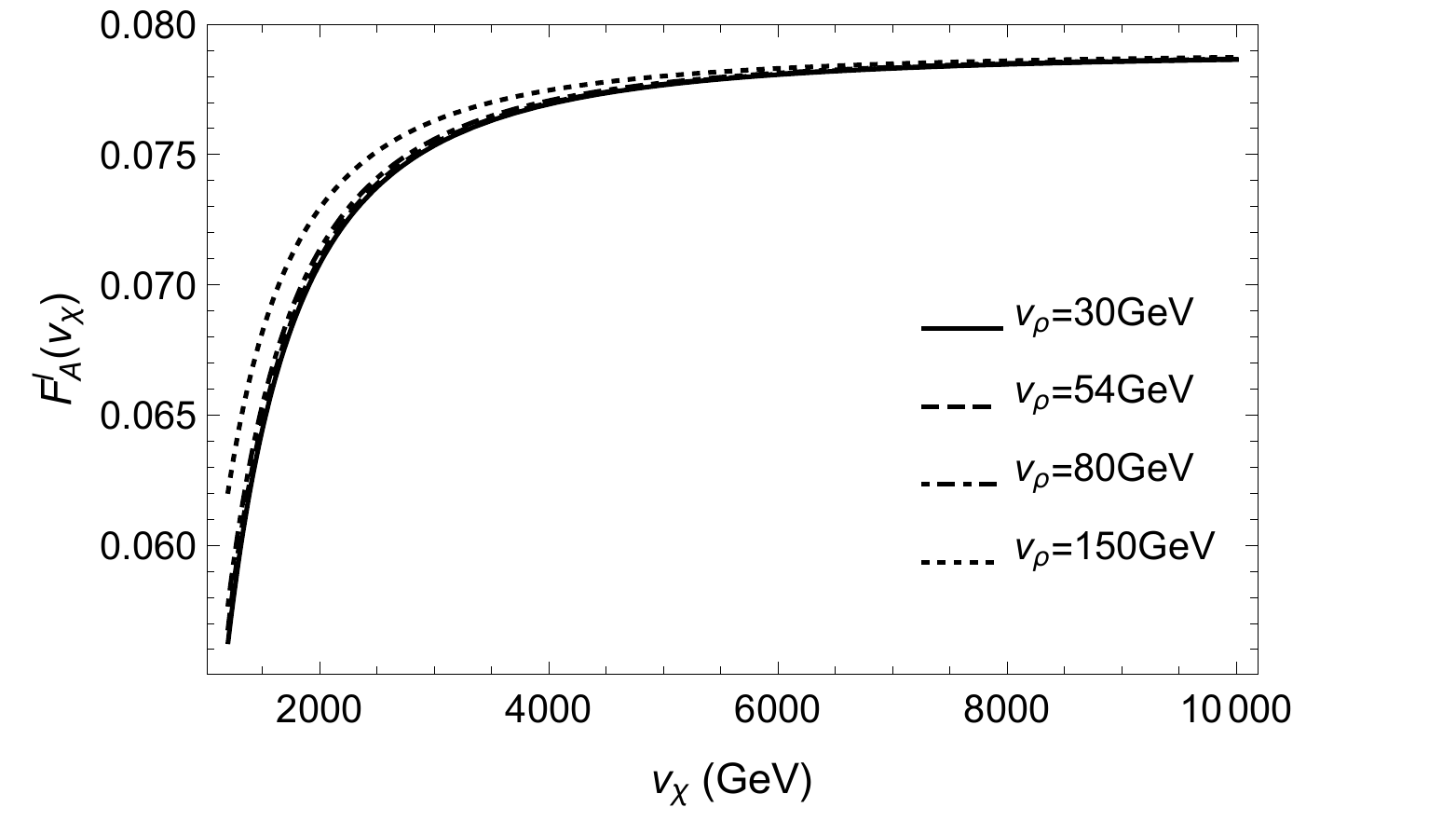}}
	\hfill
	\subfloat[\label{4c}]{
		\includegraphics[height=0.27\columnwidth]{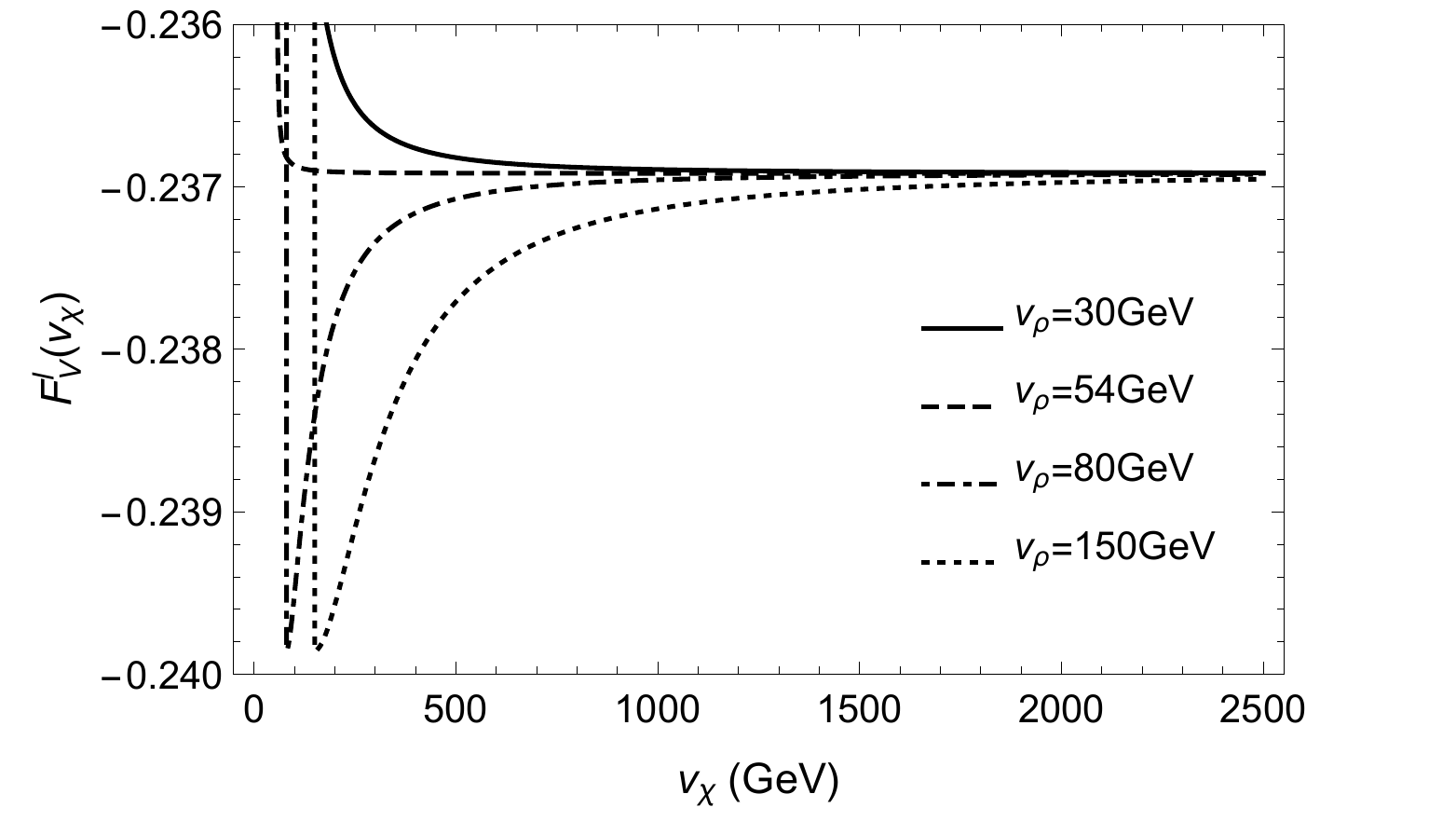}}
	\caption{ Predictions of the lepton neutral current couplings with $Z_2$ as a function of $v_\chi$ and $v_\rho$ at Z-pole. (a) neutrino couplings, $F^\nu_{V,A}$; (b) the charged lepton axial couplings, $F^l_A$; (c) the charged leptons vector coupling, $F^l_V$. We have defined $f^i_{V,A}c_W\equiv F^i_{V,A}$. }
	\label{fig:quatro}
\end{figure}

\begin{figure}[h!]
	\subfloat[\label{5a}]{\includegraphics[height=0.27\columnwidth]{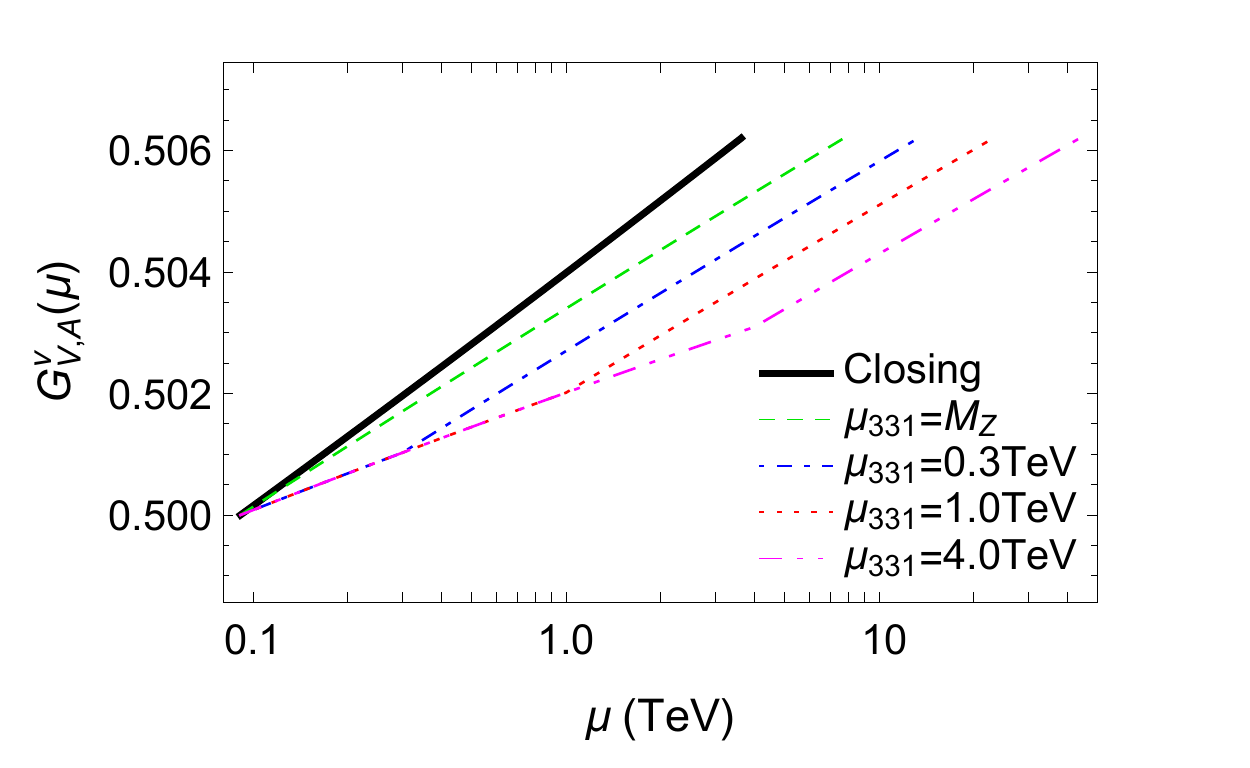}}
	\hfill
	\subfloat[\label{5b}]{\includegraphics[height=0.27\columnwidth]{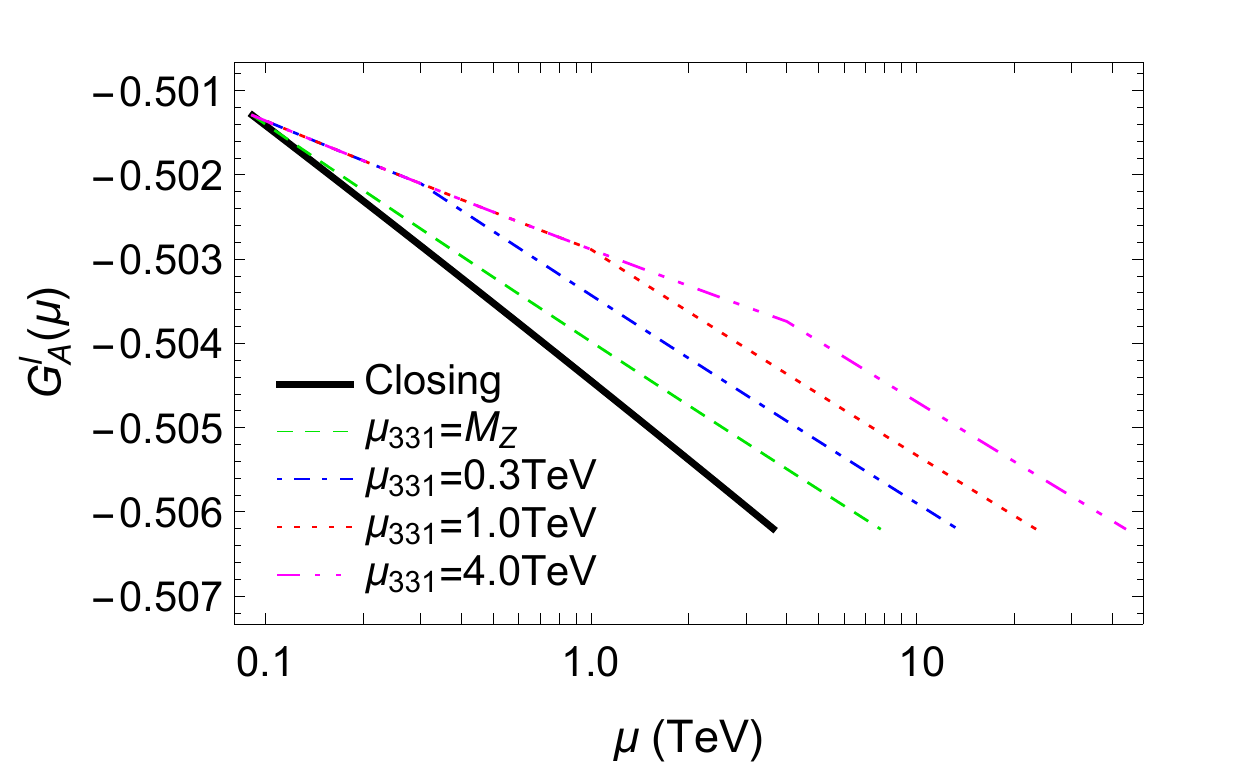}}
		\hfill
	\subfloat[\label{5c}]{\includegraphics[height=0.27\columnwidth]{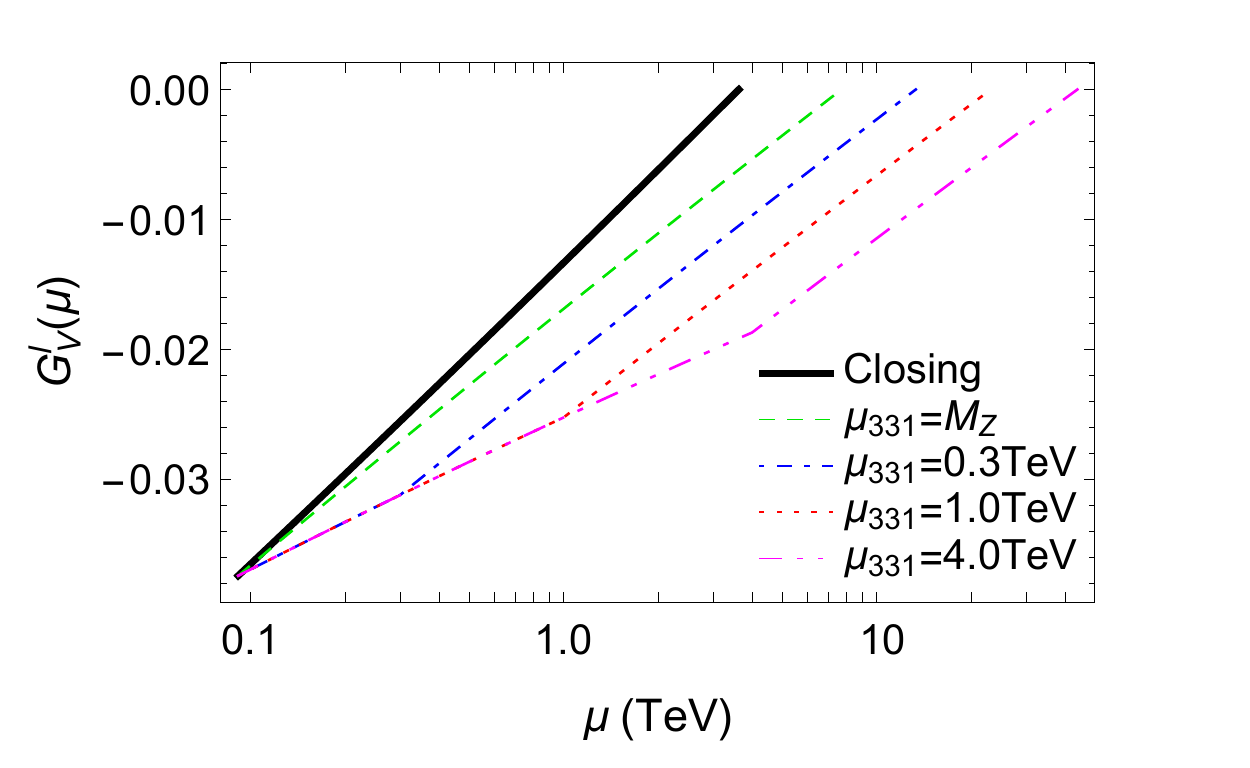}}
	\caption{Lepton neutral current couplings with $Z_1$ as a function of energy for fixed $v_\chi=2.0$ TeV and $v_\rho=54$ GeV: (a) the neutrino couplings; (b) the charged lepton axial couplings; (c) the charged leptons vector couplings.}
	\label{fig:cinco} 
 \end{figure}

\begin{figure}[h!]
	\subfloat[\label{6a}]{\includegraphics[height=0.27\columnwidth]{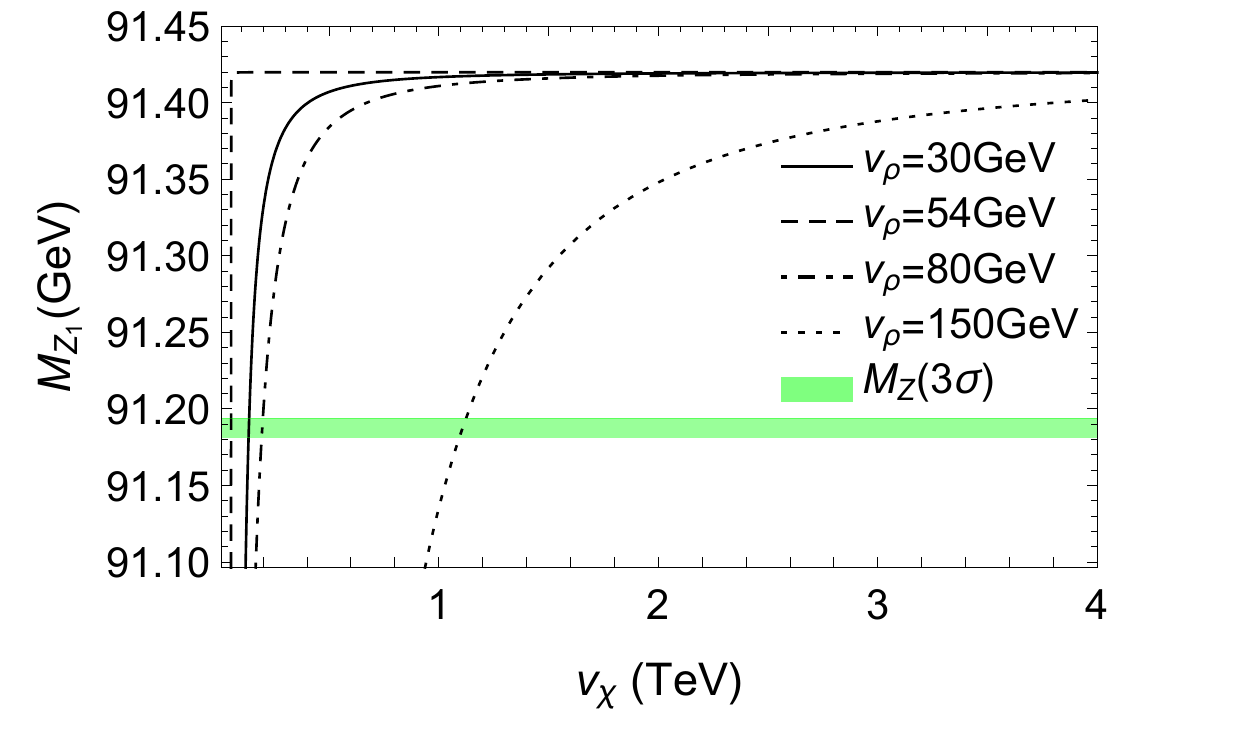}}
	\hfill
	\subfloat[\label{6b}]{\includegraphics[height=0.27\columnwidth]{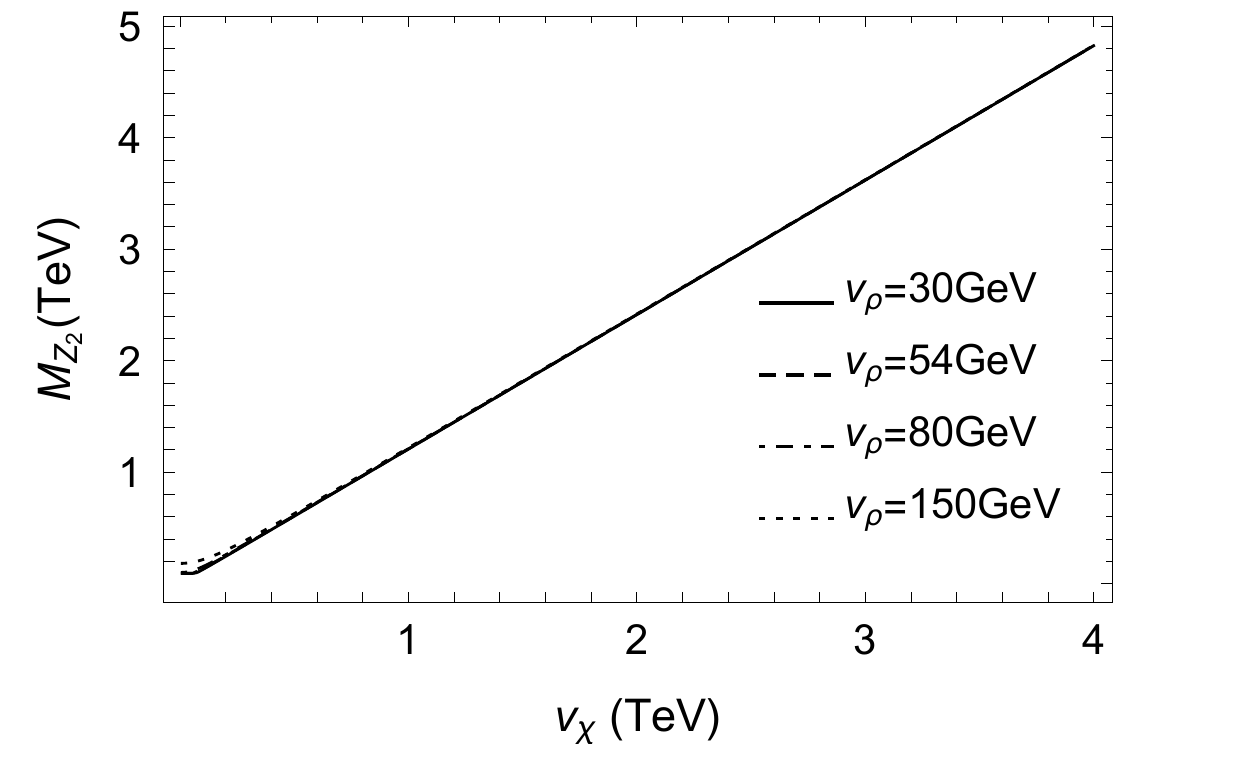}}
	\caption{Neutral vector boson masses as function of $v_\chi$ and $v_\rho$ at Z-pole: (a) curve of $M_{Z_1}$, with the 3-$\sigma$ experimental band of $M_Z$; (b) curve of $M_{Z_2}$. }
	\label{fig:seis} 
\end{figure}

From $\mu_m\leq \mu\leq \mu_{331}$ we consider an intermediate situation in which
the complete 331 symmetry is used (but not all the m331 degrees of freedom): three lepton triplets $\psi_i\sim (1,3,0)$; two quarks (anti)triplets $Q_a\sim(3,3^*,-1/3)$; one quark triplet $Q_3\sim(3,3,2/3)$; three singlets $u_{iR}\sim(3,1,2/3)$ and three singlets $d_{iR}\sim(3,1,-1/3)$;  the scalar triplets $\eta\sim (1,3,0),\rho\sim(1,3,+1)$ plus only the SM gauge bosons. The behavior of $s^2_X$ according Eq.~(\ref{run1}) is shown by the lower continuos curve in Figs.~\ref{fig:dois}. In fig.~\ref{2a} we consider only the triplet $\eta$ and in Fig.~\ref{2b} we consider both triplets, $\eta$ and $\rho$.
Considering only the $\eta$, we have ($i=U(1)_{X},SU(3)_{L},SU(3)_{C}$) :
\begin{equation}
b^{331(\textrm{reduced}-\eta)}_{i}=\left( \frac{22}{3},-\frac{19}{6},-6\right) .
\label{aa4}
\end{equation}
and with both triplets,
\begin{equation}
b^{331(\textrm{reduced}-\eta\rho)}_{i}=\left( \frac{25}{3},-3,-6\right) .
\label{aa5}
\end{equation}

For the full particle content of m331 model embedded in the full 331 symmetry, we have
\begin{equation}
b^{331}_{i}=\left( \frac{122}{9},-\frac{17}{3},-6\right) .
\label{aa6}
\end{equation}

\newpage

\end{document}